\documentclass[%
reprint,
 amsmath,amssymb,
 aps,
prab,
]{revtex4-2}

\usepackage{graphicx}
\usepackage{dcolumn}
\usepackage{bm}
\usepackage{algorithm,algcompatible}
\usepackage{textcomp}
\usepackage{siunitx}

\begin{document}

\preprint{APS/123-QED}

\title{Expanding LUME to Support Virtual Accelerators and Digital Twins}

\author{Ryan Roussel}
\author{Sara Miskovich}
\author{Gopika Bhardwaj}
\author{Jeremy Lorelli}
\author{Auralee Edelen}
\affiliation{SLAC National Accelerator Laboratory, Menlo Park, CA 94025, USA}

\author{Christopher M. Pierce}
\author{Ken Lauer}
\author{Christopher Mayes}
\affiliation{xLight, Palo Alto, CA 94306, USA}

\date{\today}

\begin{abstract}
Virtual accelerators and digital twins are increasingly essential tools for accelerator operations, controls development and verification, and model-based optimization. However, current implementations are often tightly coupled to specific simulation codes, facilities, and applications, resulting in fragmented, ad hoc solutions that are difficult to reuse or extend. 
To address this, we expand the LUME Python package to include standardized implementation and deployment of virtual accelerators and digital twins across heterogeneous simulation backends and control system interfaces. 
At the core of this change is the introduction of LUMEModel abstraction, which defines a fixed, simulator-agnostic API and a variable system that encodes metadata such as units and data types/validation. 
This design enables standardized interaction with physics-based simulators, surrogate models, and differentiable simulations, while supporting both Python-native workflows and IOC-based operation via EPICS using the lume-pva package. 
Facility- and simulator-specific details are encapsulated through extensible transformer layers, allowing consistent control-system semantics to be mapped onto diverse simulation engines. 
We describe the LUMEModel architecture, variable system, and package ecosystem, and present representative use cases including model interchangeability, staged and chained simulators, and continuous integration testing. This work will make implementing and using virtual accelerators easier and more flexible.
\end{abstract}

\maketitle




\section{Introduction}

Virtual accelerators (VAs) and digital twins (DTs) have become important components of modern accelerator facilities, supporting a wide range of activities including operations, machine development, controls software testing, and model-based optimization. By providing simulation-backed representations of accelerator behavior, these systems enable safe experimentation, rapid prototyping of control strategies, and deeper integration of beam dynamics simulations and surrogate models into day-to-day operations. As accelerator facilities grow in complexity and increasingly rely on automated and adaptive control algorithms, the need for robust, reusable, and extensible virtual accelerator infrastructures has become more pronounced.

Despite their importance, existing virtual accelerator and digital twin implementations are often developed in a facility specific manner \cite{liuzzo2022esrf}. Individual applications frequently rely on bespoke integrations between specific beam dynamics codes, local control system conventions, and narrowly defined use cases. As a result, significant duplication of effort occurs across facilities and projects, with limited reuse of software components and little standardization of interfaces. These tightly coupled designs make it difficult to interchange simulator backends, extend existing systems to new facilities, or support emerging workflows such as differentiable modeling, surrogate-based optimization, and continuous integration testing of controls software.

The Lightsource Unified Modeling Environment (LUME) \cite{mayes_lightsource_nodate} framework was previously developed to aid in unifying the execution of charged particle beam simulations by providing Python wrappers for running and modifying different simulation codes such as Impact \cite{qiang_three-dimensional_2006} and Genesis \cite{reiche_genesis_1999}. In this work, we extended the scope of LUME to address the challenges of creating virtual accelerators and digital twins by providing a standardized, modular Python framework for their creation. At the core of this approach is the \texttt{LUMEModel} class, which defines a fixed API for setting and retrieving model variables, resetting internal state, and publishing supported control variables with associated metadata. This abstraction enables heterogeneous simulators such as physics-based codes, neural network surrogates, and differentiable models to be used interchangeably within a common control and application layer.

In addition to its Python-native interface, \texttt{LUMEModel} supports operation within traditional accelerator control system environments. Through the \texttt{lume-pva} package, \texttt{LUMEModel} instances can be exposed as EPICS Process Variable (PV) servers, allowing virtual accelerators and digital twins to be accessed by existing control room tools and applications without modification. This dual-mode operation allows a single implementation of accelerator simulations to be used for multiple purposes via different APIs.

This paper presents the design and implementation of the \texttt{LUMEModel} framework, with an emphasis on its model abstraction, variable system, simulator- and facility-specific extensions, and Input/Output Controller (IOC) integration. 
Representative use cases are presented to demonstrate how \texttt{LUMEModel} enables model interchangeability, staged and chained simulation workflows, continuous integration testing, and model calibration using mixed black-box and differentiable simulators.
Together, these capabilities position \texttt{LUMEModel} as a scalable foundation for virtual accelerator and digital twin development, with a clear pathway toward future integration with higher-level modeling abstractions such as the Particle Accelerator Lattice Standard (PALS).

\section{Background}

\subsection{Definitions}

In this section, we provide working definitions for several commonly used terms in virtual accelerator / digital twin modeling. These terms often have varying meaning among practitioners and we choose to make are use clear here.

\paragraph{Virtual Accelerator}
A virtual accelerator (VA) is a representation of an accelerator or accelerator subsystem that mimics the behavior of the physical machine while remaining fully decoupled from live hardware. A virtual accelerator typically exposes the same control interfaces, variable names, and units used by the real machine, allowing applications, control software, and algorithms to interact with it without needing modifications to do so. Virtual accelerators are commonly used for software development, testing, operator training, model calibration, and design studies, and may be composed of physics-based simulations, surrogate models, or combinations thereof.

\paragraph{Digital Twin}
A digital twin (DT) is a specialized form of a virtual accelerator that is explicitly synchronized with a real accelerator through live measurements and sends information back to the machine (bi-directional information flow). In contrast to a generic virtual accelerator, a digital twin is intended to reflect the current or recent operating state of the machine, including configuration, calibration parameters, and measured conditions. Digital twins are often used for real-time decision support, online optimization, what-if analysis, and predictive diagnostics, and may run continuously alongside the physical accelerator.

\paragraph{Simulator and Surrogate Model}
A simulator refers to a physics-based computational model that evaluates accelerator behavior based on first-principles or reduced-physics descriptions, such as particle tracking, field solvers, or beam envelope equations. Examples include Bmad \cite{sagan_bmad_2006}, IMPACT \cite{qiang_three-dimensional_2006}, Cheetah \cite{kaiser_bridging_2024}, or similar beam dynamics codes. A surrogate model, by contrast, is an approximate model constructed to emulate the behavior of a simulator or physical system, often using machine learning or statistical techniques. Surrogate models are typically optimized for speed or differentiability and may trade physical interpretability for computational efficiency. Both simulators and surrogate models are treated uniformly within \texttt{LUMEModel} through a common abstraction layer.

\paragraph{IOCs and Process Variables}
In accelerator control systems, an IOC is a server process that manages communication between control applications and hardware or software endpoints. An IOC exposes named PVs, which represent readable and/or writable quantities such as device setpoints and measurements. 
Within LUME, virtual accelerators and digital twins can be exposed as IOCs, with PVs derived directly from model variables, enabling integration with existing control room tools and workflows.


\subsection{Applications of Virtual Accelerators and Digital Twins}

Virtual accelerators and digital twins enable a broad spectrum of applications across accelerator operations, modeling, and software development. At LCLS and LCLS-II, these tools have become increasingly important as machine complexity and availability demands continue to grow. 
While the specific application areas differ, they share a common requirement: a realistic, controllable representation of accelerator behavior that can be accessed through standard control system interfaces.

\paragraph{Virtual Diagnostics}
At LCLS, virtual accelerators are used to provide synthetic diagnostics corresponding to physical measurements such as beam profiles from wire scans and screen images. For example, in the LCLS CU-HXR beamline, simulated screens and beam moments generated by physics-based models or surrogate neural networks can be exposed as virtual diagnostic PVs. 
These diagnostics can provide live information to accelerator operators about the beam distribution at locations within the accelerator when measurements would otherwise be invasive or destructive.

\paragraph{Online Control and Decision Support}
Digital twins at LCLS can be initialized from archived or live machine states to provide predictive decision support during operations. In this mode, operators and applications can evaluate the impact of proposed configuration changes such as magnet strengths or RF parameters using a synchronized digital twin before applying them to hardware. This capability is particularly valuable during machine development shifts and high repetition rate operations, where rapid iteration and risk mitigation are critical.

\paragraph{Model Calibration}
Virtual accelerators can be used at LCLS to calibrate accelerator models against experimental data. For instance, in LCLS-II diagnostic sections such as DIAG0, measured screen images and beam parameters can be compared against predictions from beam dynamics codes or differentiable surrogate models embedded within a virtual accelerator. These workflows enable continuous refinement of model accuracy, supporting both offline studies and online applications.

\paragraph{Controls Software Testing and Continuous Integration}
A major application of virtual accelerators at LCLS is controls software testing, especially as accelerator control tends towards automation. For example, a virtual accelerator exposed via EPICS IOCs, is used to perform nightly testing of software used by Badger \cite{zhang_badger_2022} to interact with the real accelerator. This enables regression testing, validation of new software releases, and long-term maintenance of controls infrastructure, independent of machine availability and without impacting operations.

\paragraph{Operator and Algorithm Training}
Virtual accelerators also serve as training platforms for both human operators and automated control algorithms at LCLS. Operators can rehearse standard tuning procedures and response to off-normal conditions using realistic machine models, while optimization and machine learning algorithms can be trained and evaluated on virtualized accelerator instances. This reduces risk to hardware and accelerates the deployment of advanced control techniques.


\subsection{Limitations of the Current State of Practice}

Despite the promise of virtual accelerators and digital twins, their development and deployment have historically followed fragmented and facility-specific paths. These limitations arise not from fundamental shortcomings in simulation capabilities, but from the way models, controls systems, and applications have traditionally been integrated.

A central challenge is the highly fragmented ecosystem of beam dynamics codes and simulators. Accelerator modeling relies on a diverse set of physics-based tools, each with its own execution model, configuration syntax, data formats, and assumptions. Codes such as Bmad and IMPACT are typically optimized for specific classes of problems, making it difficult to combine them within a single workflow or to interchange simulators without significant re-engineering of surrounding software. Similarly, the introduction of machine learning based surrogate models adds an additional interface to the mix that must be integrated with conventional beam dynamics simulations.

Integration is further complicated by the prevalence of facility-specific details that must be accounted for when transitioning from a beam dynamics simulation to a virtual accelerator that is emblematic of the real accelerator interface. 
To bridge the gap between simulation codes and operational software, many projects develop bespoke glue code tightly tailored to a single accelerator, application, or study. These wrappers often encode implicit knowledge about control names, units, coordinate systems, and simulation behavior, making them difficult to generalize, reuse, or maintain over time.

Collectively, these factors create significant barriers to reuse, extensibility, and systematic validation. Software developed for one facility or application is rarely portable to another without extensive modification. Testing often depends on access to the physical machine or specialized environments, limiting the adoption of continuous integration and regression testing practices. As accelerator facilities increasingly depend on software-driven operations and advanced modeling techniques, these limitations motivate the need for a more standardized, modular approach to virtual accelerator and digital twin implementation.


\section{LUME Model}

Enhancements to the LUME framework are designed to address the limitations of existing virtual accelerator and digital twin implementations by introducing a small set of clear, principled standards for APIs.

\begin{figure}
    \includegraphics[width=\linewidth]{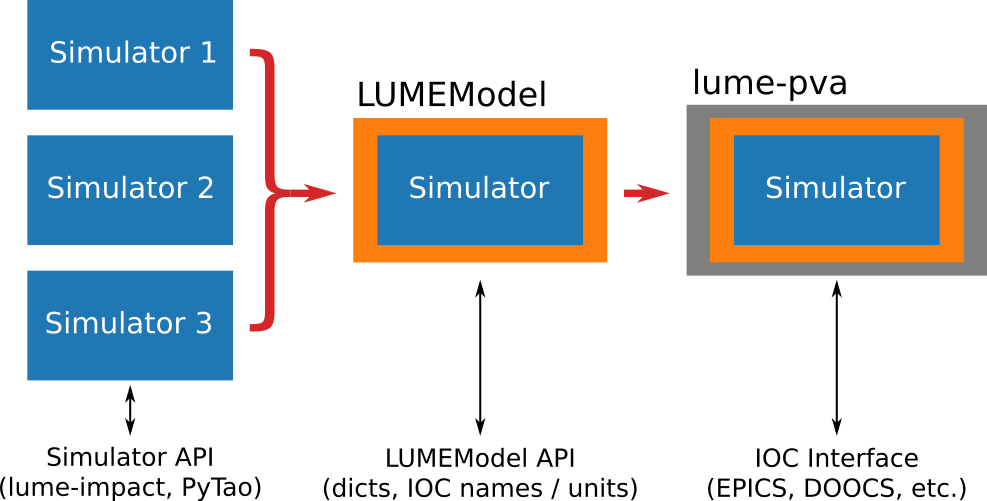}
    \caption{Overview of \texttt{LUMEModel} design philosophy for standardizing API's for interacting with different simulator backends such as LUME-Impact, PyTorch, etc.}
    \label{fig:overview}
\end{figure}

To this end, we introduced the \texttt{LUMEModel} class inside LUME to provide a standardized Python API for interacting with accelerator simulators as virtual accelerators and digital twins. Rather than exposing simulator-specific interfaces to downstream applications, \texttt{LUMEModel} objects define a fixed model abstraction through which parameters are set, outputs are retrieved, and model state is managed. This approach enables applications to interact with accelerator models in a uniform manner, independent of the underlying simulation engine.

LUME is also designed to support both Python-native and IOC-based workflows. The same \texttt{LUMEModel} object can be accessed directly through a Python API for scripting, analysis, and algorithm development, or exposed as a control system server using EPICS via the \texttt{lume-pva} package.
This dual-mode operation enables reuse of the same underlying models across offline studies and operational deployment.

Here we describe the core classes that are used in conjunction with \texttt{LUMEModel}.

\subsection{Core Architectural Components}

The \texttt{LUMEModel} class defines a fixed Python interface through which all accelerator models are accessed, regardless of their internal implementation. This interface specifies methods for setting input variables, retrieving output variables, resetting internal state, and advertising the set of supported control variables along with their associated metadata. By enforcing this common contract, LUME ensures that applications and control interfaces remain agnostic to the details of the underlying simulator.

\begin{figure*}
    \includegraphics[width=\linewidth]{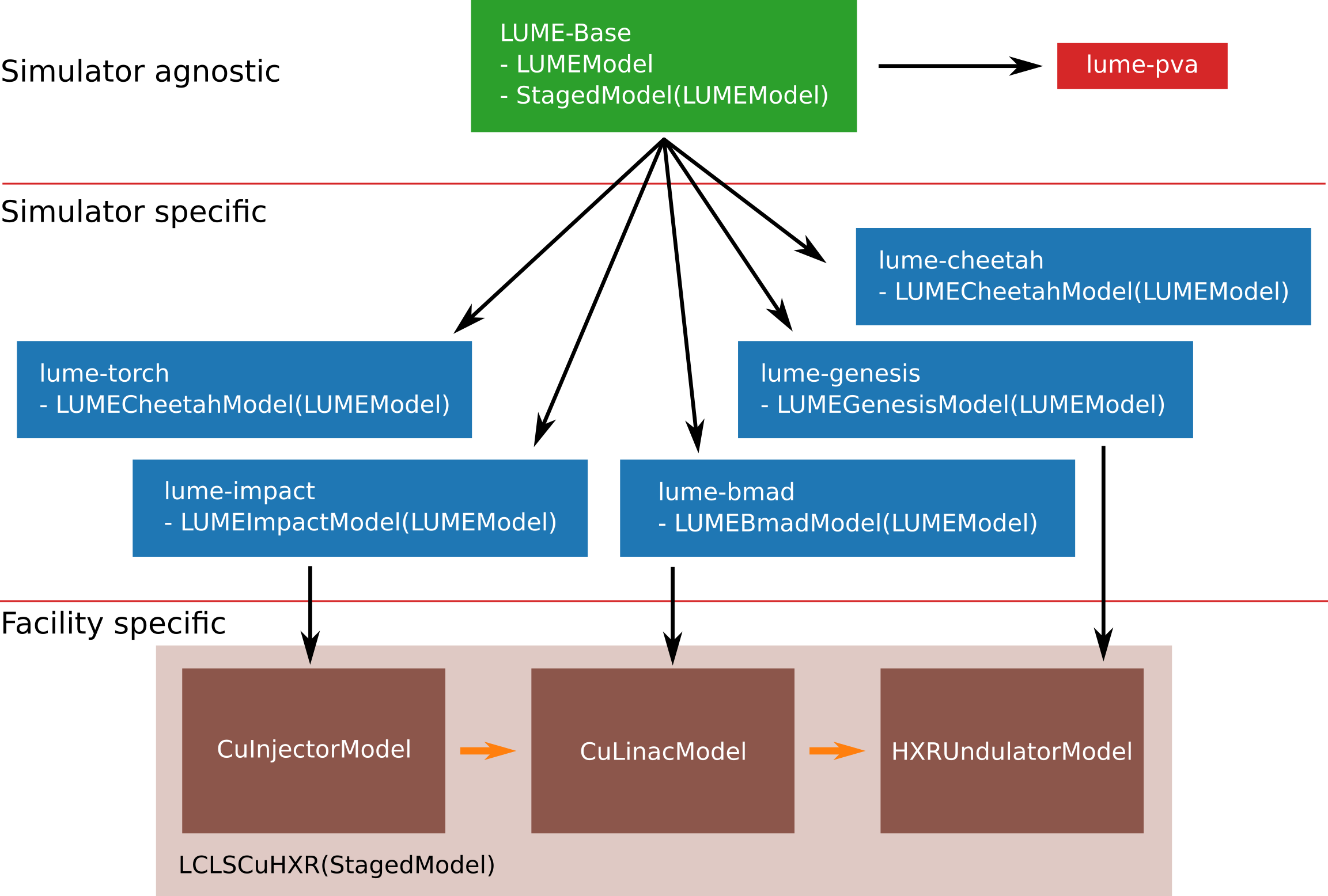}
    \caption{Overview of LUME packages and modules for virtual accelerator and digital twin implementations. The \texttt{LUMEModel} class serves as an abstract class for simulator agnostic interactions with a virtual accelerator. Simulator-specific packages define subclasses which implement the needed abstract methods. Facility-specific packages implement necessary logic to utilize simulator-specific models to create a virtual accelerator. Models from heterogenous LUME packages can be used in a single virtual accelerator, providing a unified interface that is independent of internal model types. All models defined as a subclass of \texttt{LUMEModel} can be used to create EPICS PVs via \texttt{lume-pva}.}
    \label{fig:packages}
\end{figure*}

Built on top of the \texttt{LUMEModel} abstraction are a collection of simulator-specific LUME packages that can be used to assist in creating model objects, as shown in Fig.~\ref{fig:packages}. These packages encapsulate the logic required to interact with simulators such as beam dynamics codes or surrogate models. Each backend is responsible for translating between the generic \texttt{LUMEModel} API and the simulator’s native configuration, execution, and data extraction mechanisms. This design allows multiple simulators to be supported concurrently and enables rapid interchangeability of modeling approaches without changes to downstream applications. On the other hand, users can implement custom behavior by directly subclassing the abstract model class.

Control system integration is provided through the \texttt{lume-pva} package, which uses the metadata defined by LUME variables to automatically generate PVs with appropriate names, units, access permissions, and data types. Simulation execution and PV handling are coordinated within a multithreaded runtime, allowing models to respond to control system requests in a manner consistent with operational IOCs. Through \texttt{lume-pva}, the same LUME model used in Python-based workflows can be deployed directly within existing control system environments.




\subsection{LUMEModel Class Description}

The \texttt{LUMEModel} abstract class specifies three primary methods for interacting with model state. Together, these methods define a complete interaction cycle for configuration, execution, and data retrieval.

The \texttt{get()} method retrieves the current values of one or more model variables, identified by their control-system-facing names. This method provides read access to simulated diagnostics, internal state variables, or computed outputs, and returns values in forms consistent with their associated metadata, such as data type and shape.

The \texttt{set()} method updates model inputs by assigning new values to one or more variables. These variables typically correspond to control parameters such as magnet strengths, RF phases, or model configuration flags. The setter interface allows multiple variables to be updated in a single call, supporting atomic configuration changes and efficient application-level coordination.

The \texttt{reset()} method returns the model to a defined baseline state. This may include resetting cached simulation results, restoring default parameters, or reinitializing internal simulator objects. The reset operation is particularly important for reproducibility, testing, and batch execution workflows, where model state persistence must be carefully controlled.

In addition to these methods, each \texttt{LUMEModel} instance exposes a \texttt{supported\_variables} property. This interface describes the set of variables supported by the model and provides associated metadata such as units, read/write permissions, data dimensionality, and variable type. This metadata is used by applications and control system integrations to validate interactions and to automatically construct user-facing interfaces.

Notably, to handle backend input/output computations, \texttt{LUMEModel} instances store an instance of the underlying simulator class within the modeling instance. This allows Python users to interact directly with the backend simulator via its own API, enabling hybrid manipulations of the simulator, both via the virtual accelerator \texttt{set} and \texttt{get} methods and the simulator API.




\subsection{LUME Variable System}

The LUME variable system provides the mechanism by which accelerator model inputs and outputs are validated and exposed to applications and control systems. Variables describe the names, data types, and read/write access that an instance of  \texttt{LUMEModel} supports. By encapsulating this metadata, the variable system enables consistent interaction, automatic validation, and integration with control-system semantics via \texttt{lume-pva}.

Input validation is performed when variables are set through the \texttt{set()} interface, allowing improper values, incompatible shapes, or unauthorized write attempts to be detected early. Similarly, output validation during \texttt{get()} ensures that simulator results match the declared variable definitions before being exposed to users or control systems.

Metadata stored within variables also plays a central role in control system integration. Units and access permissions are used to define appropriate PV properties when models are exposed as IOCs, while shape and type information informs client applications how to interpret returned data. By centralizing this information within the variable layer, LUME avoids duplication of control-system logic across simulators and facilities.

\subsubsection{Variable Types}

Within the LUME framework, several core variable types are defined to support the range of data commonly encountered in accelerator modeling and diagnostics.

\paragraph{ScalarVariable}
The \texttt{ScalarVariable} type represents float-like scalar quantities, such as magnet strengths, RF phases, or single-valued diagnostic signals. Scalar variables are typically used for control parameters and summary outputs and are defined with associated units and read/write permissions. Subclasses of the \texttt{ScalarVariable} type specify specific sub-types including ints, floats, and booleans.

\paragraph{NDVariable}
The \texttt{NDVariable} type represents multi-dimensional, array-like data. Common examples include screen images, wire-scan profiles, and multi-channel waveform data. In addition to units and access permissions, NDVariables specify expected dimensionality and shape, enabling consistent handling of structured diagnostic data across simulators and control interfaces.

\paragraph{ParticleGroupVariable}
The \texttt{ParticleGroupVariable} type is used to represent particle ensemble data, leveraging the \textit{openPMD-beamphysics} \texttt{ParticleGroup} data structure. This variable type supports rich phase-space representations, including particle coordinates, momenta, and weights. ParticleGroupVariables enable advanced applications such as phase space reconstruction, beam distribution analysis, and hybrid workflows that combine particle-based simulations with surrogate or differentiable models.

\paragraph{Torch Variable Extensions}
For \texttt{LUMEModel} subclasses that utilize PyTorch computations internally, we define PyTorch-specific variable types in the \textit{lume-torch} package, \texttt{TorchScalarVariable} and \texttt{TorchNDVariable}, that use PyTorch tensor objects instead of floats and arrays to store inputs and outputs. This enables backpropagation through \texttt{LUMEModel} instances, allowing for fast gradient calculations on CPUs and GPUs that can be used in gradient-based optimization algorithms.


\section{Simulator-Specific and Facility-Specific Implementations}
To optionally reduce the burden on users, we created simulator-specific subclasses of \texttt{LUMEModel} which implement boilerplate code needed to interact with each simulator. This structure allows simulator-specific logic to evolve independently while maintaining a consistent external interface. These packages can be found at \texttt{https://github.com/lume-science}.

\paragraph{lume-impact}
\texttt{lume-impact} provides an implementation of the \texttt{LUMEModel} interface for IMPACT-T / IMPACT-Z \cite{PhysRevSTAB.9.044204, qiang1999object} based beam dynamics simulations. It encapsulates the translation between LUME variables and IMPACT input decks, execution control, and output parsing.

\paragraph{lume-bmad}
\texttt{lume-bmad} interfaces LUME with the Bmad \cite{sagan_bmad_2006} accelerator simulation library through PyTao. It enables accelerator lattices and configurations defined in Bmad to be exposed via the standardized \texttt{LUMEModel} API. Facility-specific subclasses and transformer layers map control-system variables to Tao commands, making \texttt{lume-bmad} a core component for virtual accelerators and digital twins at facilities such as LCLS.

\paragraph{lume-cheetah}
\texttt{lume-cheetah} integrates the Cheetah \cite{kaiser_bridging_2024} modeling framework into the LUME ecosystem. Cheetah provides fast, ML-oriented beam dynamics models optimized for GPU acceleration and surrogate-based workflows. Through \texttt{lume-cheetah}, these models can be combined with traditional physics-based simulators or deployed independently in applications requiring rapid evaluation or batched execution.

\paragraph{lume-torch}
\texttt{lume-torch} extends the LUME variable and model system to support PyTorch-based, fast-executing surrogate models.

\paragraph{Facility-Specific Model Packages}
Facility-specific models are built on top of simulator-specific LUME packages to encode local control conventions, accelerator configurations, and deployment requirements. Here \texttt{LUMEModel} subclasses for particular machines or beamlines (e.g.\ LCLS or LCLS-II) and include transformer logic that maps facility control variables to simulator parameters. An example of facility specific virtual accelerator implementations using LUME can be found at \texttt{https://github.com/slaclab/virtual-accelerator}.

\section{Staged Model Composition}
Staged models in LUME enable the composition of multiple simulator instances into a single logical accelerator model. In this approach, an accelerator is decomposed into sequential stages, each represented by its own \texttt{LUMEModel} backend. These stages may correspond to physical beamline segments, such as an injector, linac, or undulator section, or to functional subdivisions chosen to optimize simulation performance and accuracy. Particle distributions are passed from one stage to the next via standardized openpmd-beamphysics \cite{mayes_christophermayesopenpmd-beamphysics_2026} \texttt{ParticleGroup} objects which are exposed as \texttt{\detokenize{initial_particles}} and \texttt{\detokenize{final_particles}} properties added to \texttt{LUMEModel} subclasses via \texttt{InitialParticlesMixIn} and \texttt{FinalParticlesMixIn} mix-in classes.

From the perspective of applications and control systems, a staged model behaves as a single \texttt{LUMEModel} instance. LUME manages the propagation of variables and data between stages, coordinating execution order and ensuring consistency of shared quantities such as beam distributions. This allows complex accelerator configurations to be represented in a modular fashion while preserving a unified external interface.

The LCLS CuHXR beamline provides a representative example of how LUME supports advanced model composition by combining both staged modeling and heterogeneous simulation modalities within a single, unified virtual accelerator. This approach reflects practical operational needs at LCLS, where different sections of the accelerator demand different trade-offs between computational speed, physical fidelity, and mathematical structure.

In the CuHXR implementation, the beamline is decomposed into sequential stages that correspond to physically and functionally distinct accelerator subsystems, such as the injector, linac, and downstream transport and diagnostic sections. Each stage is implemented as its own \texttt{LUMEModel} object, selected to best match the modeling requirements of that segment. LUME coordinates the execution of these stages and manages the propagation of shared state—most notably beam distribution data—such that, from the perspective of applications and control systems, the staged chain behaves as a single logical accelerator model.

The CuHXR staged model takes advantage of heterogeneous modeling approaches to improve VA performance depending on the application. For example, the LCLS injector can be modeled by either a detailed beam dynamics simulation implemented in Impact or using a surrogate model of the injector trained on simulation and experiment data. Since \texttt{LUMEModel} provides a standardized interface for both, it is easy to switch between models of the injector as needed in the staged model, depending on whether or not greater speed or accuracy is needed by the specific application. The interchangeability of model types is available down the accelerator modeling chain as new beam dynamics models or surrogates are developed.

\section{EPICS and IOC Integration with lume-pva}

A key capability of the LUME framework is the ability to deploy virtual accelerators and digital twins directly within operational control system environments. This functionality is provided by the \texttt{lume-pva} package, which enables \texttt{LUMEModel} subclasses to be exposed as IOCs using EPICS PVs. Through this integration, models implemented using LUME can be accessed by existing control room applications and tools without modification.

\texttt{lume-pva} supports the automatic creation of EPICS PV servers from \texttt{LUMEModel} instances. By interrogating the model’s \texttt{supported\_variables} interface, \texttt{lume-pva} dynamically generates PVs corresponding to each exposed model variable. This process eliminates the need for manual IOC creation and ensures that the control system representation remains consistent with the underlying model definition.
Additionally, \texttt{lume-pva} provides a client mode of operation that allows it to consume remote PVs which in-turn, updates the internal \texttt{LUMEModel} instance.

The behavior of each PV is defined using the metadata provided by the LUME variable system. Units, read/write permissions, data types, and shapes specified in variable definitions are propagated directly into the EPICS layer. As a result, PVs exposed by \texttt{lume-pva} closely mirror those of physical hardware, supporting realistic interaction patterns for both operators and automated applications. Input validation performed at the variable level is preserved when PVs are written, preventing invalid or unauthorized updates from reaching the simulation backend.

To support responsive operation and scalable deployment, \texttt{lume-pva} employs a multithreaded execution model that separates PV handling from simulation execution. Incoming PV updates are collected asynchronously, while simulation runs are scheduled and executed in a coordinated manner. This design allows virtual accelerators to service control system requests efficiently while managing potentially expensive simulation workloads, and supports deployment scenarios ranging from lightweight local testing to containerized digital twins.

Although \texttt{lume-pva} currently targets EPICS-based control systems, its design is intentionally extensible. The separation between model logic, variable metadata, and IOC runtime enables alternative control system backends to be implemented with minimal changes to core LUME components. This provides a clear pathway for future integration with other control frameworks, such as DOOCS \cite{hensler1996doocs}, and aligns with LUME’s broader goal of supporting evolving control system infrastructures while preserving a stable model interface.

\section{Deployment Modes}

\begin{figure*}
    \includegraphics[width=\linewidth]{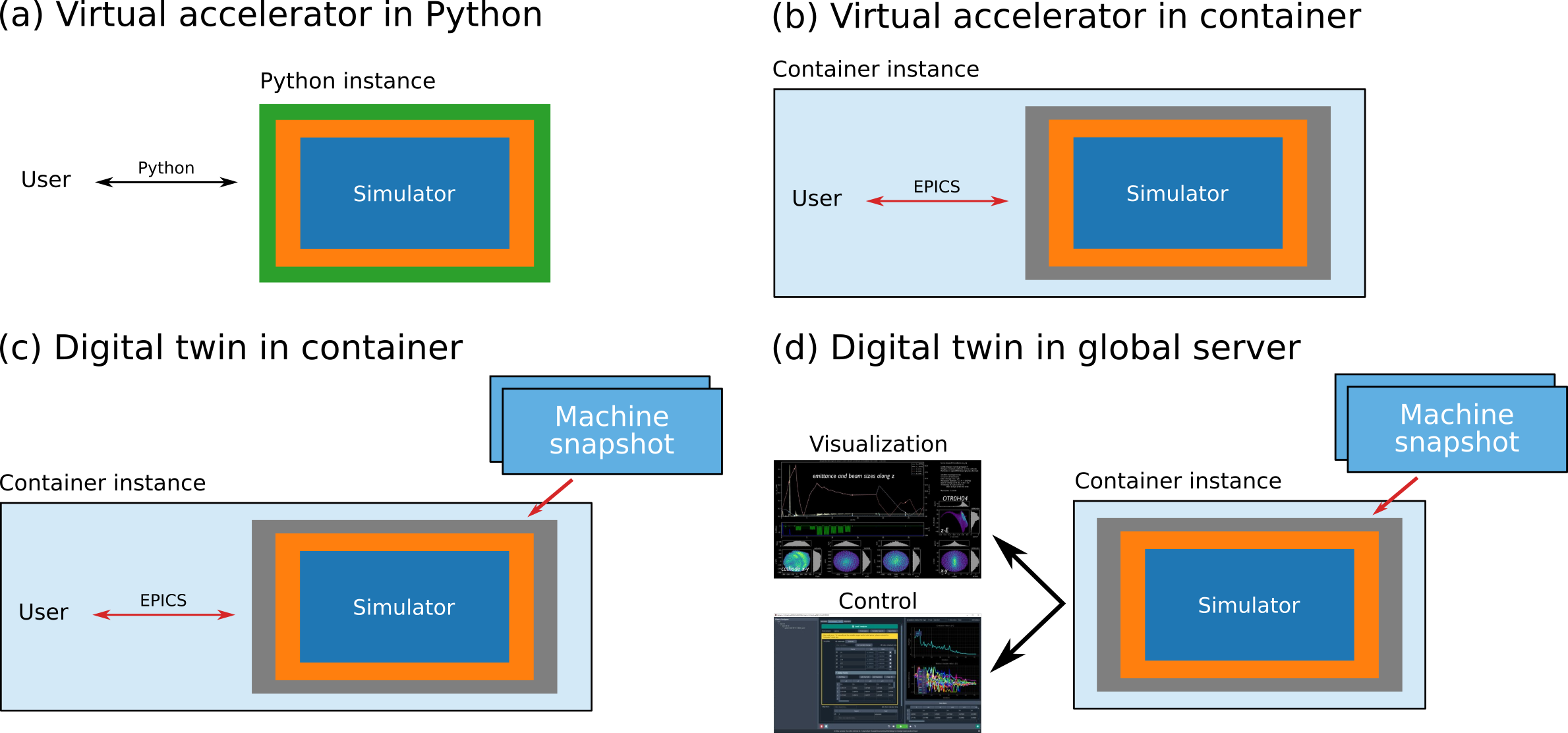}
    \caption{Virtual accelerator and digital twin deployment modalities using \texttt{LUMEModel}. (a) \texttt{LUMEModel} object defined inside a Python instance, enabling direct interaction with the underlying simulation API. (b) \texttt{LUMEModel} instance wrapped by \texttt{lume-pva} to serve EPICS PVs inside a container. (c) Digital twin in a container which enables EPICS interactivity with updates from the live / historical machine state. (d) Global digital twin server that constantly updates with the machine state and provides read-only access to model PVs.}
    \label{fig:deployment}
\end{figure*}

The LUME framework is designed to support a range of deployment modes as shown in Fig.~\ref{fig:deployment} that reflect the varying requirements of accelerator modeling, software development, and operations. Here we highlight a few interaction modes that enable applications of interest inside and outside the accelerator control room.

\paragraph{Local Python-Based Virtual Accelerators}
In the simplest deployment mode, LUME-based virtual accelerators are instantiated and exercised directly within a Python environment. This mode is well suited for algorithm development, data analysis, and exploratory modeling, where tight integration with scientific Python workflows is desirable. Users interact with the \texttt{LUMEModel} API directly, enabling rapid iteration and easy integration with optimization libraries, machine learning frameworks, and custom analysis code. Local Python-based deployments are also commonly used in unit tests and developer-level validation of simulator backends. Finally, interacting with LUME-based virtual accelerators inside a Python environment enables access to the underlying simulator object API, enabling flexibility in controlling the virtual accelerator behavior outside the \texttt{set()} and \texttt{get()} methods.

\paragraph{Containerized Virtual Accelerators}
For improved reproducibility and portability, LUME virtual accelerators can be deployed within containerized environments. Containerization encapsulates simulator dependencies, Python environments, and model configurations, allowing consistent behavior across development, testing, and deployment platforms. At facilities such as LCLS, containerized virtual accelerators are used to support automated testing pipelines and shared development environments, ensuring that models behave identically regardless of host system configuration. Additionally, interacting with EPICS within an independent container prevents conflicts with real machine parameters and states, preventing errant machine commands / readbacks from effecting real accelerator operations without needing to change the naming convention of EPICS PVs for the virtual accelerator.

\paragraph{Containerized Digital Twin Connected to Live Machines}
LUME also supports deployment of containerized digital shadow that are initialized from, and periodically synchronized with, live machine states. Users of the digital shadow can also manually trigger model synchronization to the live machine or historical operating states. This deployment mode enables predictive evaluation of configuration changes, online diagnostics, and modeling-assisted operations while isolating simulation execution from the physical control system infrastructure.

\paragraph{Centralized Digital Twin Servers}
In larger-scale deployments, multiple LUME-based digital twins can be hosted on centralized servers and accessed by a broad set of users and applications via read-only interactions. These centralized services may support multiple concurrent instances, persistent machine snapshots, and shared visualization or monitoring tools. At LCLS and similar facilities, such deployments enable coordinated access to virtual accelerators for operations staff, machine physicists, and software developers, while maintaining consistent model definitions and control interfaces. Centralized digital twin servers provide a natural integration point for visualization frameworks and future orchestration layers built on higher-level control abstractions.

\section{Example Applications}
The design of the LUME framework enables a range of practical use cases that arise in accelerator modeling, controls development, and operations. This section highlights representative examples, with emphasis on applications drawn from LCLS and LCLS-II.

\paragraph{Online Virtual Diagnostics and Visualization}
Virtual accelerators and digital shadows enable real-time visualization of beam dynamics and virtual diagnostics that are difficult, destructive, or impossible to observe directly during accelerator operations. By embedding a virtual accelerator within a control application, or by hosting a digital shadow on a local or remotely accessible server, model outputs, such as beam phase spaces and Twiss parameters, can be visualized either instantaneously or tracked continuously over extended periods of operation. These capabilities provide operators with deeper insight into machine behavior, supporting informed control decisions and accelerating online troubleshooting.

\paragraph{Prior Models for Bayesian Optimization}
Previous demonstrations \cite{boltz_leveraging_2025} have shown that incorporating prior models of objective functions into the Gaussian Process model used to make control decisions in Bayesian optimization algorithms substantially increases convergence speed to optimal operating configurations. By using a standardized API, digital twins provided by \texttt{LUMEModel} can be used for this purpose inside the Xopt \cite{roussel_xopt_2023} optimization library, regardless of backend simulator type.

\paragraph{Continuous Integration and Software Testing}
A major operational use case at LCLS is the deployment of virtual accelerators for controls software testing. One such example of this is testing the code used in Badger \cite{zhang_badger_2022} to communicate with the machine control system and execute measurements. To do this, LUME-based virtual accelerators are exposed via \texttt{lume-pva} as EPICS IOCs and used in automated nightly test workflows to validate functionality, detect regressions, and ensure compatibility with evolving machine configurations. This approach enables continuous integration practices that would be impractical if testing were limited to the physical accelerator.

\paragraph{Model Calibration and Phase Space Reconstruction}
In LCLS-II diagnostic sections, measured data from screens and other diagnostics can be compared against predictions from virtual accelerators built with LUME to calibrate models of the machine, similar to \cite{kaiser_bridging_2024, gonzalez-aguilera:ipac23-wepa065, gupta_improving_2021} and reconstruct incoming beam distributions \cite{PhysRevAccelBeams.27.094601}. Both black-box physics simulations and differentiable surrogate models can be embedded within the same framework, enabling hybrid optimization workflows. These capabilities facilitate systematic parameter fitting, uncertainty analysis, and reconstruction of beam distributions under realistic operational conditions.


\section{Future Directions}

The LUME framework is intended to evolve alongside accelerator modeling and control system infrastructure. While its current design addresses immediate needs in virtual accelerator and digital twin development, several future directions have been identified to improve scalability, interoperability, and long-term sustainability.

A primary focus of future development is closer alignment with PALS-based infrastructure. PALS aims to provide a standardized, high-level description of accelerator components, control variables, and operational semantics across facilities. By integrating PALS metadata as an upstream source of configuration and control information, future LUME implementations can reduce duplication of facility-specific logic while preserving the existing \texttt{LUMEModel} abstraction. This alignment would allow simulator backends to be selected and configured dynamically based on PALS descriptions, further decoupling model logic from control system details.

Closely related to this goal is the reduction of facility-specific coupling. Current LUME deployments rely on facility-specific transformer layers to map local control conventions to simulator parameters. While this approach enables practical deployment today, it introduces maintenance overhead and limits portability. Future work will focus on minimizing bespoke mappings by leveraging shared metadata standards and common transformation patterns, allowing virtual accelerators to be more easily transferred between facilities with similar accelerator structures.

Another important direction is the continued standardization of transformers and metadata. As additional simulator backends and facility implementations are developed, common patterns are emerging in how control variables are translated into simulator configurations and how outputs are interpreted. Formalizing these patterns into reusable transformer interfaces and metadata schemas will improve consistency across LUME packages and reduce the effort required to support new simulators or beamlines.

Finally, the evolution of LUME is expected to be increasingly community-driven. As the framework is adopted across multiple facilities and projects, feedback from users, developers, and operations staff will guide refinements to the API, variable system, and deployment tooling. By maintaining a clear separation between stable core abstractions and extensible implementation layers, LUME can accommodate new modeling approaches and control paradigms while preserving backward compatibility for existing applications. Additionally, we wish to develop LUME such that it is interoperable with other packages for virtual accelerator and digital twin implementation \cite{brynes2026flexible, brynes2026closing}.

\section{Conclusions}

This work has presented updates to the LUME framework that support implementing virtual accelerators and digital twins in a standardized, extensible manner. 
By introducing the \texttt{LUMEModel} abstraction, a variable system with explicit metadata, LUME addresses many of the limitations present in existing ad hoc implementations. 
The framework enables heterogeneous modeling approaches, including physics-based simulators, surrogate models, and differentiable simulations, to be accessed through a common interface and deployed consistently across development and operational environments.

A central benefit of LUME is its emphasis on modularity and standardization. Simulator-specific details, facility-dependent control conventions, and control system interfaces are encapsulated within clearly defined layers, reducing coupling and duplication of effort. This structure allows accelerator models to be reused across applications, facilitates rapid interchange of simulator backends, and supports staged and composite models that reflect the complexity of modern accelerators. The ability to deploy the same models through both Python-native workflows and EPICS-based IOCs further enhances flexibility and promotes consistency across use cases.

Artificial intelligence tools (i.e., text-to-text generative AI) were used in the preparation of a draft of this report and in editing. The authors take full responsibility for the contents of this manuscript.

\bibliography{LUME-model}

\end{document}